\documentclass{bmcart}

\usepackage{amsthm,amsmath}
\usepackage[utf8]{inputenc} 

\startlocaldefs
\usepackage{array}
\usepackage{booktabs}
\usepackage{multirow}
\usepackage{changepage}
\usepackage{makecell}
\usepackage{graphicx}
\endlocaldefs

\begin{document}

\begin{frontmatter}

\begin{fmbox}
\dochead{Research}


\title{Exploiting potentialities for space-based quantum communication network: downlink quantum key distribution modelling and scheduling analysis}


\author[
addressref={aff1,aff2},                   
]{\inits{X. W.}\fnm{Xingyu} \snm{Wang}}
\author[
addressref={aff2},
]{\inits{C. D.}\fnm{Chen} \snm{Dong}}
\author[
addressref={aff1},
]{\inits{J. W.}\fnm{Jiahua} \snm{Wei}}
\author[
addressref={aff2},
]{\inits{T. W.}\fnm{Tianyi} \snm{Wu}}
\author[
addressref={aff1},
]{\inits{T. L.}\fnm{Taoyong} \snm{Li}}
\author[
addressref={aff1},
]{\inits{H. Y.}\fnm{Huicun} \snm{Yu}}
\author[
addressref={aff1},
]{\inits{S. Z.}\fnm{Shanghong} \snm{Zhao}}
\author[
corref={aff1},                       
email={slfly2012@163.com}   
]{\inits{L. S.}\fnm{Lei} \snm{Shi}}


\address[id=aff1]{
	\orgdiv{School of Information and Navigation},             
	\orgname{Air Force Engineering University},          
	\city{Xi'an},                              
	\cny{China}                                    
}
\address[id=aff2]{%
	\orgdiv{Information and Communication College},
	\orgname{National University of Defense Technology},
	\city{Xi'an},
	\cny{China}
}



\end{fmbox}

\begin{abstractbox}

\begin{abstract} 
	With the goal of a space-based quantum network is to have satellites distribute keys between any nodes on the ground, we consider an evolved quantum network from a near-term form, in which a space-based relay, satellite executes a sequence of satellite-based quantum key distribution (SatQKD) missions, allowing any two ground nodes to have a shared secure key. Accordingly, we develop a comprehensive framework for the dynamic simulation of SatQKD and consider scheduling QKD downlink in future space-based quantum communication network. The embedding of precise orbital model including beam diffractions and atmosphere effects makes our framework more realistic. Incorporated with the local meteorological data to channel loss model, the cloud cover contribution to transmission disturbance is quantified to realize a quasi-experimental scenario. We provide a trade-off consideration between the duration available for communications and its corresponding average link loss during one orbit, which could be used to support decisions involving the locations of ground stations and the selection of the orbital parameters for a quantum satellite. Our work also shows that satellite downlink schedule could allow for the possibility to consider strategies for different SatQKD missions such as extending connection for distant ground nodes, prioritized delivery, or promoting keys utilization, which can be used as a guideline to support decisions for future satellite application.
\end{abstract}




\end{abstractbox}
%

\end{frontmatter}



\section*{INTRODUTION}
Quantum key distribution (QKD) is a family of protocols that can provide information-theoretic security to share keys between two distant parties~\cite{RevModPhys.74.145,PhysRevLett.68.557,10.1145/382780.382781}. In addition to fibre-based QKD approaches, free-space QKD has progressed out of laboratories into real-world scenarios~\cite{PhysRevLett.81.3283,PhysRevLett.94.150501,Ursin2007,PhysRevLett.98.010504}. Several experiments~\cite{Yin2012,Liao2017,Wang2017,Yin2017,Yin2020}, such as the Chinese Quantum Experiments at Space Scale (QUESS) missions contributing to full in-orbit demonstrations of satellite-based QKD (SatQKD) approaches, provide a feasible option for achieving an ultralong-distance QKD relying on a direct transmission. To date, SatQKD has already shown its own distinct advantages not only to overcome the distance limit of ground-based links, but also to connect intercontinental nodes, even to be integrated with fibre links to join different networks~\cite{Chen2021}.

An example of the integrated space-to-ground network illustrated in Fig.~\ref{Fig1illustration}, where ground stations act as hubs that connect to ground-based network using fibre-based trusted nodes, is most likely feasible form in the near term. Since the fibre-based trusted nodes are in fixed locations that could be subject to constant surveillance and probes~\cite{Diamanti2016}, the ultimate vision of a satellite quantum communication network is to have satellites distribute keys to ground nodes. Inspired by practical implementations, theoretical proposals aimed to be pathfinders for a network have been made that employ downlinks, inter-satellite links or others. Remarkably, by building polar-orbit constellations of low-Earth orbit (LEO) satellites, that continually perform SatQKD with the optical ground station (OGS) they fly over, it is possible to offer symmetric keys between any two OGSs~\cite{Vergoossen2020,WangJY2021,Mazzarella2020,Sidhu2021,Huang2020,Grillo2021}.

To date, satellites have been recognized as either the trusted node or untrusted node~\cite{Mustafa2021}, for which the quantum satellite could directly mediate the distribution of secure encryption keys pairwise between the nodes. The remarkable fact that, the ‘untrusted node’ configuration such as the implemented entanglement-based QKD~\cite{Khatri2021} or future space-based measurement-device-independent (MDI) QKD~\cite{PhysRevLett.125.260503} is more secure but feasible only when both OGSs are within the satellite coverage simultaneously~\cite{PhysRevX.9.041012,Wang_2021}. This leads to a very short time window for communication with OGSs. While in the ‘trusted node’ mode, the satellite carries out QKD operations with distinct OGSs to establish independent keys with each of them and subsequently broadcasts the XOR hash of both delivered keys over a public channel, thus allowing any two ground nodes to have a shared key~\cite{PhysRevLett.120.030501}. This scheme allows encrypted communication in the ground-based network to be implemented without the fibre-based relays.

However, one major challenging bottleneck in putting SatQKD into extensive use is that some current physical and technological limitations in the satellite scenario~\cite{Bedington2017,Kaltenbaek2021,Singh2021}. To overcome this issue, one can one can either upgrade the QKD system to achieve a faster data collection rate or spend more time to gather data. As for the former, improving the experimental settings (e.g., employing larger lens) is regarded as beneficial for a lower downlink channel loss, while from a practical engineering perspective, a trade-off between cost and efficiency to determine space-qualified parameters exists~\cite{Bedington2016,Oi2017,Manuel2021}. The latter, a state-of-the-art QKD system installed in a geosynchronous orbit (GEO) satellite~\cite{Vallone2016,Dirks2021} is considered to operate 24-hour QKD, while the yields of long-term SatQKD in the daytime are unclear owing to the unpredict solar radiation~\cite{LiaoSK2017}. Moreover, in a near term, satellites only have one source on board~\cite{Villar:20}, which can perform SatQKD to only one OGS at a time. These limitations hence suggest that, to improve the SatQKD efficiency, an efficient scheduling of downlink to the set of OGSs is required.

Investigations of observation duration scheduling problems for Earth observation satellite are being used to execute highly efficient Earth observation tasks. Inspired by the similar satellite-to-ground orbit operations~\cite{Mustafa2019,Kim2021}, the authors of ref.~\cite{Polnik2020} elegantly extended the idea of scheduling downlinks to the scenario of satellite-to-ground QKD. To design a sequence of assignment that the satellite executes, ultimately achieving the optimal schedule, it is typically evaluated in terms of secure keys available. This requirement makes it necessary to have a prior comparison of the expected results of all possible satellite operations. For all of results, the unique characteristics of optical transmission from satellite to ground should be considered. In addition, the communication window for OGSs also have variable durations that depend on various factors in the satellite scenario. Taken together, atmospheric conditions, transmission path lengths and orbital propagations should be considered in a comprehensive, dynamic simulation of SatQKD.

In this work, we develop a comprehensive modelling framework of SatQKD. We consider orbital propagations to determine orbital parameters at each time step, the time-varied satellite-to-ground channel loss including beam diffractions and atmosphere effects can be as a function of propagation distance and elevation angle. Moreover, Incorporated with the local meteorological data to loss model, the cloud cover contribution to transmission disturbance is quantified, thus a recreation of the quasi-experimental scenario. Choosing the Micius satellite experimental setting for paradigms of satellite applications, we simulate a course of Micius satellite flying across the ground stations built on the specific locations of Chinese nodes, and estimated the maximum number of keys that is possible to deliver to the ground nodes during a specific week. Furthermore, we listed the satellite-to-ground downlink optical transmission loss for all ground nodes, and capture its variations by selection of the orbital altitudes for a satellite. Meanwhile, we provide a trade-off consideration between the duration available for communications and its corresponding average link loss during an orbit, which could be used to support decisions involving the locations of ground stations and the selection of the orbital parameters for a satellite. Lastly, we demonstrate that the use of schedule is an effective way in networking a system of OGSs. Exploiting potentialities of satellite downlink schedule allows for the possibility to consider strategies for different SatQKD missions such as extending connection for distant nodes, prioritized delivery, or promoting keys utilization.

\section*{METHODS}

In this section, we will first introduce the SatQKD co-modelling framework. Here, orbital propagations and downlink SatQKD simulations are two key components in the co-modelling framework. Then, this lays the groundwork for scheduling optimization by providing the SatQKD keys results under all possible schedule cases.

\subsection*{SatQKD co-modelling framework}

\subsubsection*{Scenario modelling with AGI Systems Tool Kit software}

The time-varying scenario constraints due to satellite movement can be predicted with orbital dynamics. As shown in Fig.~\ref{Fig2framework}, the ephemeris data of Micius, the two-line elements (TLE) are collected to perform orbital propagations and to cover orbit drifts. Supposed that the latitude-longitude-altitude (LLA) of the involved ground stations, the visibility including relative position and the elevation of the Micius satellite from each ground station, at each time step, can typically be assessed with SGP4 algorithm. For the convenience of calculation, our framework is integrated with AGI System Toolkit's modelling capabilities, we can obtain a specific scenario concerning visibility, relative elevation angle and distance from a satellite to the destination, ultimately implementing the modelling of the dynamic communication links with discretization.

Here, the intervals available for communications can be defined as the times when the satellite is in the darkness of night time and the elevation angle between the satellite and a node is greater than 10 degrees. Thus, the elevation angle and the relative propagation distance from the satellite to a given node will be passed to be used as an input for the subsequent link loss modelling.

\subsubsection*{Modelling loss as function of transmission length and evaluation angle}

Optical power signal launched from the transmitter is affected by various factors until it is finally detected at ground-based receiver. The downlink channel loss depends on the atmospheric conditions, such as turbulence and weather conditions~\cite{Dequal2021} and on the scenario parameters, such as propagation distance and elevation angle, in which geometric loss and atmospheric loss contribute significantly. The prediction of the geometric loss owing to the beam takes the divergence angle, the link distance, and the receiver lens aperture size as inputs, and it scales as the inverse square of the propagation distance, with the beam width typically being several times larger than the diameter of the receiving telescope. The geometric loss $\Lambda_{\rm{G}}$ (dB) can be expressed as~\cite{Khalighi2014}

\begin{equation}
\Lambda_{\rm{G}} = 1 - \exp \left( { - \frac{{{D_r}^2}}{{2{\omega _r}^2}}} \right)
\end{equation}
\begin{equation}
\omega_r = \sqrt {{{\left( {\frac{\lambda }{{\pi \varphi }}} \right)}^2} + {{\left( {\varphi L} \right)}^2}} 
\end{equation}

Here, $L$ is the propagation distance between the OGS and the satellite, $D_r$ is the receiver diameter, $\omega_r$ and $\lambda$ is the received beam width and the wavelength, respectively. $\varphi$ represents the diffraction limited and atmospheric turbulence induced divergence angle. In general, the definition of diffraction limited divergence angle is a function of the transmitter diameter $D_t$, which is calculated as $1.22\lambda/{D_t}$. However, we do not want to underestimate the effect of atmospheric turbulence, we refer the results that Micius, with $D_t$ of 300 mm, reported 10 $\rm{\mu rad}$ divergence angle. Indeed, estimating the effect of transmitter diameter $D_t$ is more complex since factors other than diffraction, such as pointing performance, also determine the time-averaged ground spot size. Consider that the optical beam propagates through atmosphere in the downlink, beam spreading leads to a less significant pointing loss $\Lambda_{\rm{P}}$ compared to the uplink, thus we consider it as a fixed value.

On the other hand, loss occurs due to atmosphere effects, e.g., absorption and scattering scales exponentially with propagation distance as a result of Beer-Lambert law. Though the atmospheric effects are relevant only in a layer of thickness 10–20 km above the earth’s surface, the practical propagation distance is a straight line along the elevation angle, which will lead to path elongation effects at low elevation angle. This can be elevated as ${\Lambda _{\rm{A}}} = {\Lambda _{{\rm{A,0}}}}/\sin\theta$, where $\theta$ is elevation angle, and $\Lambda _{{\rm{A,0}}}$ is determined by the signal wavelength, e. g., 2 dB for 1550 nm~\cite{Kerstel2018}. Therefore, using a series of parameters associated with the specifications of Micius satellite, the losses including the beam diffractions and atmosphere effects can be regarded as a function of propagation distance and elevation angle.

\subsubsection*{Using meteorological data to quantify cloud cover disturbance}

Clear skies in the two localized areas are typically assumed in the modelling. As a result, the variability in the real-world results is missed, and the days with cloud disturbance are excluded from the execution of the QKD missions. Followed the method proposed in ref.~\cite{Bonato2009,Pirandola2021}, we here considered the meteorological data from Himawari-8 historical cloud statistics, in which the cloud optical thickness for each location coordinate was encoded as an integer value varied from 0 to 150.

Note that the data set updates every 10 minutes, and thus cloud contribution to transmission can be assumed as 

\begin{equation}
{\rm{ }}{\Lambda _{\rm{C}}} = {\rm{ - 10lo}}{{\rm{g}}_{10}}\frac{{150 - \alpha }}{{150}}{\rm{, 0}} \le \alpha  < 150
\end{equation}

Here, $\alpha$ is the integer value, where the cloudless is represented as 0, while overcast corresponds to 150. The same formula for the loss of the transfer rate due to cloud cover was used by ref.~\cite{Capelle2018,Stephen2004} who studied the problem of selecting the network of OGSs. As such, the value $\Lambda _{\rm{C}}$ corresponding to the OGS locations are then assumed as extra channel loss in modelling. For convenience of analysis, we replicate the specific experimental values (see Table 2) reported in ref.~\cite{Chen2021}. Based on the effects we consider, the link loss within the available intervals separately for each time step will be exported for subsequent detection events estimation.

\subsubsection*{Estimation of the transferred secure keys}

We estimate the number of secure keys that could be obtained based on the detection events over an available interval. Here, the polarization-encoded decoy-state Bennett \& Brassard 1984 QKD (BB84) for the implementation of SatQKD is considered. Using Gottesman–Lo–Lütkenhaus–Preskill (GLLP) security analysis, the asymptotic key rate in the case is given by~\cite{Ma2005,Lo2005}

\begin{equation}
R_{{\rm{GLLP}}} = q\left\{ { - {f_e}\left( {{E_\mu }} \right){Q_\mu }{h_2}\left( {{E_\mu }} \right) + {Q_1}\left[ {1 - {h_2}\left( {{e_1}} \right)} \right]} \right\}
\end{equation}

where $q$ depends on the implementation (1/2 for BB84 protocol because half of the time, Alice and Bob disagree with the bases, and if one uses the efficient BB84 protocol, $q\approx 1$), $f_e$ is the error correction inefficiency function, $\mu$ is the intensity of the signal state and  $h_2$ is the binary entropy function.   and $Q_1$ are the gain and $e_1$ error rate of the single photon states estimated using decoy-state theory, respectively. The gain $Q_\mu$ and quantum bit error rate (QBER) $E_\mu$ for all the photon number components is estimated by using the standard channel model, in which the transmittance can be obtained by

\begin{equation}
\eta = {\rm{ 1}}{{\rm{0}}^{ - \frac{{{\Lambda _{\rm{C}}} + {\Lambda _{\rm{A}}} + {\Lambda _{\rm{G}}} + {\Lambda _{\rm{P}}}}}{{10}}}}
\end{equation}

Therefore, the total number of secure keys over the interval from a start time $a$ to an end time $b$ can be calculated as the sum of the rates per second

\begin{equation}
K_{a,b} = \sum\limits_{i = a}^b {{R_{GLLP}}[\eta ({t_i})]}
\end{equation}

where $\eta ({t_i})$ is the transmittance at time $t_i$.

\subsection*{Scheduling downlink for one satellite to multiple stations}

\subsubsection*{Problem formulation}

The problem of scheduling a QKD downlink from a single satellite to multiple ground stations was studied by ~\cite{Polnik2020}. The work showed that a Mixed-Integer Program (MIP) with time discretisation can be used to solve to optimality for problem instances based on the expected key results of all possible satellite operations. Different from previous work, our work here aims to schedule one satellite to all possible ground stations in the network allocating them time suitable to generate keys, ultimately achieve a distribution of keys that allows for the possibility to consider strategies for different missions such as extending connection for distant nodes, prioritized delivery, or promoting keys utilization.

Before formally introducing the problem, we first repeat some fundamental definitions that would be used. Followed the method of time discretization, let the simulation period $T$ be divided into intervals with a same length of 10 s. Here, we denote the number of intervals as $M$. Thus, $K_m^n$ represents the number of secure keys that can be sent to nodes ${\rm{ }}n \in [1,....,N]$ during an arbitrary interval ${\rm{ }}m \in [1,....,M]$. Note that, all physical constraints are handled in the definition of $K_m^n$. For instance, in the interval $m$ when the OGS located in node $n$ is in the darkness of night time or the elevation angle between the satellite and the OGS is smaller than 10 degrees, $K_m^n$ will return 0. Moreover, considering the time spent switching between consecutive deliveries from one node to another, we add an interval to represent the switch before the next access, which can be expressed as

\begin{equation}
\left\{ {\begin{array}{*{20}{c}}
	{\sum\limits_{n \in N} {\sum\limits_{m \in M} {x_m^n \le 1} } }\\
	{x_{m + 1}^n \le x_m^n + x_m^0{\rm{  }}}
\end{array}} \right.
\forall n \in [1,...,N],\forall m \in [1,...,M]
\end{equation}

Here, the binary variable $x_m^n$ describes whether the interval $m$ is assigned to node $n$. In constraint (7), the first inequality implies that at every time in the period, at most one node can be assigned. The second inequality ensures that the required switching period $x_m^0$ before a handoff to other nodes is considered. That is, if interval $m+1$ is assigned to a node, then either the preceding interval $m$ should be scheduled for the same transmission or the requested switch $x_m^0$ should be performed at interval $m$.

With the above constraints, we now formulate the optimization problem. Our goal in optimization is to find the optimal schedule that maximizes the total number of secure keys under the different strategies involving (1) general delivery to the nodes without considering weights, pursing only the maximization of the total final keys (denoted as S-GD); (2) prioritized delivery to nodes with higher weights, ensuring that high-priority tasks are completed first (denoted as S-PD); (3) targeted delivery to nodes with respect to distinct weights, making proportions coincident with the network traffic distribution (denoted as S-TD). 

For this, a general problem formulation for these strategies can be formulated as
\begin{equation}
\begin{array}{l}
{\rm{Max}} \sum\limits_{n = 1}^N {E_n^T} \\
{\rm{Subject to}} E_n^T = \sum\limits_{m = 1}^M {K_m^nx_m^n} \\
{\rm{                        }}\sum\limits_{n = 1}^N {\sum\limits_{m = 1}^M {x_m^n \le 1} } {\rm{                  }}\\
{\rm{                         }}x_{m + 1}^n \le x_m^n + x_m^0{\rm{      }}\forall n \in [1,...,N],\forall m \in [1,...,M]
\end{array}
\end{equation}

where ${\rm{ }}E_n^T{\rm{ }}$ is the number of secure keys delivered to node $n$ over the simulation period $T$, which is product of the number of secure keys and the resulting binary variable ${\rm{ }}x_m^n$.

To obtain optimal solutions in different strategies, we performed the optimization based on the genetic algorithm (GA). In general, exploration of the search space is through by the crossover- and mutation-driven recombination of solutions, whereas the fitness-based selection ensures the property of convergence. In this case, we start with a population of randomly generated individuals. For S-GD, we consider the cost function to be only sum of the secure keys, which is calculated as ${\rm{ Fitness(}}I{\rm{) = }}\sum\nolimits_{n = 1}^N {E_n^T}$. Thus, we solve the problem by finding the maxima (i.e., the maximum fitness levels). For S-PD, the function is ${\rm{ Fitness(}}I{\rm{) = }}\sum\nolimits_{n = 1}^N {E_n^T{w_n}}$, where $w_n$ is a weight assigned to node $n$. Similarly, this allows the iterative process to tend towards higher-priority nodes when searching for a higher fitness. For S-TD, in addition to finding higher-fitness individuals by the same cost function used in S-GD, we compare the Kullback-Leibler (KL) divergence of the individuals after every fitness-based selection, which is calculated by~\cite{Barz2019}
\begin{equation}
{D_{KL}} = \sum\limits_{n = 1}^N {p({x_n})\log \left[ {\frac{{p({x_n})}}{{q({x_n})}}} \right]} 
\end{equation}

Here, $N$ is the number of nodes, $p({x_n})$ is the proportion of the number of final keys delivered to a node to all that of all nodes. $q({x_n})$ is the value of the normalized weight assigned to node $n$. That is, if two distributions match perfectly, ${D_{KL}}{\rm{ = 0}}$; otherwise, this variable takes values between 0 and $\infty$. The lower the KL divergence value is, the better matched the distribution of the delivered keys is to the traffic distribution.

Therefore, the relatively lower-value solutions are more likely to be chosen as a subset of the new population. As such, with different modifications of the GA, solving the problems of maximizing the total number of keys under different strategies are achieved. Code snippets of the above methods are illustrated in Fig.~\ref{Fig3Algorithms}.

\section*{RESULTS}

In this section, we will first demonstrate a course of the Micius satellite flying across the nodes in China by using our co-modelling framework. We simulate the satellite operations under the specific scenario associated with the TLE dataset during the week of 19th Sep. 2016 Universal Time Coordinated (UTC). Then, we will simulate the satellite-to-ground downlink transmission loss for all the ground nodes, and provide a trade-off consideration between the duration available for the communication and its corresponding average link loss during one orbit. For all of simulation results, we borrow some experimental parameters from Micius experiment, which is listed in Table.~\ref{Tab1}. Finally, we compare the distributions of keys delivered under different schedules, and consider the satellite with altitudes from LEO to GEO to evaluate the satellite schedule that yields the possible performance of the network.

First, the duration distribution of any light-of-sight ground station is depicted in Fig.~\ref{Fig4Duration}. As can be seen, the intervals are separated into two or three durations in the umbra due to the sun-synchronous orbit cycle. Consistently, within a total of 30 minutes available for performing SatQKD per day, nodes dispersed in western China, such as Lhasa and Urumqi, could immediately access the Micius satellite since they seldom shared their visible windows with other nodes. For nodes distributed in eastern China, preadoption of an appropriate schedule could achieve a reduction the time spent on the handoff resulting from random access.

Although the available intervals can be predictable with orbital dynamics, days with severe disturbance resulting from cloud cover are often excluded from the execution of QKD missions. Here, we showed an example of the cloud cover over these cities on the day of 23rd, September, 2016. Notably, as shown in Fig.~\ref{Fig5Cloudcover}(a), the cities of Xi’an and Chengdu are suffered from a serious obstruction. By incorporating the local meteorological data into our channel loss model, the cloud-induced attenuation can be quantified. As shown in Fig.~\ref{Fig5Cloudcover}(b), the total link loss in the worst case reached 37.6 dB and 33.8 dB, of which cloud disturbance accounts for approximately 4.7 dB and 2.1 dB, respectively. This extra loss will be detrimental to the efficiency in receiving the photons.

From the results in Fig.~\ref{Fig6Gloss}, we also find that the variation on link loss is mainly determined by the propagation distance, since the beam diffraction is the dominant source of loss. To further show it, we adopt the planned parameters, where the divergence angle of the satellite-based transmitting telescope is assumed to be a set of fixed values of 1, 3, 5 and 10 $\rm{\mu rad}$ and the diameter of the ground-based receiving telescope is assumed as 1.2 m. It is shown that the trend of the geometric loss is as a function of the satellite altitude for several values of divergence angle.

As can be observed, the growth rate of geometric loss for a lower divergence angle, e.g., 1 $\rm{\mu rad}$ is slower compared to that of higher divergence angles. Moreover, if the beam width is smaller than the diameter of the receiving telescope, the geometric loss is negligible, since the percentage of the received photons with respect to the transmitted one can be considered as 1. This indicates that using a larger aperture of transmitter telescope to decrease the diffraction-limited induced divergence angle could effectively improve SatQKD. However, some current in-orbit load limits are a serious concern, as is radiation damage of the optics, which make implementation of this strategy challenging in practice.

Subsequently, we explore the possibility of using a satellite with different orbits to perform schedule in a future network. we include the following orbit types for Micius in the simulation: (1) the initial orbit of the Micius satellite, i.e., an altitude of 500 km, denoted as LEO; (2) orbit altitudes of approximately 2,500 km and 5,000 km, denoted as MEO1 and MEO2, respectively; and (3) orbit altitudes of approximately 35,863 km, i.e., GEO, where the Right Ascension of Ascending Node (RAAN) of the satellite is as 50.0591 degrees for full coverage of all the nodes. For all orbits, other orbital parameters are consistent with those of the Micius satellite experiments.

Here, the duration for communication in the implementations of SatQKD is first investigated, in which the results for the simulation week are exhibited in Table~\ref{Tab2}. We can see that for the GEO satellite, even only considering their local night time, the total duration of all the ground stations is increased to about $3,587\times{10^5}$ s (99 h 25 mins), which is longer than that at LEO (204 mins), MEO1 (12 h 36 mins), and MEO2 (22 h 45 mins). This implies that one can spend more time to gather data to increase SatQKD working time with higher orbital altitudes.

On the other hand, given the link loss consideration, we also list the downlink optical transmission loss for all ground locations used in our simulation for the mentioned divergence angles. Also, for a comparison, other experimental parameters are unchanged. For all the divergence angles we consider, the link loss of SatQKD downlink from the GEO satellite to all OGSs is hard to generate a secure key rate. This indicates that scheduling appears to be difficult when the transmissions from the GEO satellite to these nodes face severe geometric losses, since few feasible alternatives exist even within a longer duration. Moreover, choosing a higher-altitude orbit satellite can effectively increase the duration but also introduces a greater loss because of the increase in the received beam width. For divergence angles of 3, 5, and 10 $\rm{\mu rad}$, the link loss of the MEO2 satellite is generally lower by one and two orders of magnitude than those of the MEO1 or the LEO satellite, respectively. In other words, since orbital period, and thus pass time, increases linearly with orbital altitude, whereas link loss increases with the square of the orbital altitude. These results imply that satellites in lower orbits deliver more key than satellites in higher orbits.

After taken the unique characteristics of the optical transmission from satellite to ground into account, we now estimate the number of secure keys for all grounds during the week. For all simulations, the sending intensities keeps a same value in ref.~\cite{Chen2021}, where $\mu {\rm{ = 0}}{\rm{.5}}$, $\nu {\rm{ = 0}}{\rm{.08}}$ and $\omega {\rm{ = 0}}$, respectively. The source repetition rate is set as 200 MHz, and other parameters are listed in Table 1. The maximum number of keys delivered to the nodes on different days of the week is shown in Fig.~\ref{Fig7keys}. We can see that the keys that could be delivered every day almost a same order of magnitude for all ground nodes. As such, from the performance of expected keys perspective, all the ground nodes have same equal ability to obtain keys. Therefore, for the ground-based communications network, the schedule, that performs a distribution of keys should be made under the desired activity of a given node in the system.

Additionally, as a demonstration, the two OGSs, i.e., Xi’an and Chengdu node at the day of 24th Sep 2016 have no keys, whereas thin clouds over the location, Wuhan can promote more keys to be delivered within the same duration from 16:32:40 UTC to 16:35:00 UTC, which indicates that the situation where clouds prevent the delivery of keys can be mitigated through a downlink schedule.

In conclusion, not only can our co-modelling framework be used as a function to formulate the scheduling problems but also these assessments related to the satellite-to-ground optical transmissions could be used to support decisions involving the locations of ground stations and the selection of the orbital parameters.

The previous section implies that using a downlink schedule can the network improves SatQKD systems efficiency, which can make the scarce SatQKD keys allocated efficiently. Here, we showed that the schedules even allow for the possibility to consider strategies for different SatQKD missions, such as the connection extension, prioritized delivery, and keys utilization promotion. 

To show their distinct advantages, let’s make some assumptions in the network. Suppose that keys downloaded from a satellite can first be stored in a buffer of a OGS and then used for postprocessing after enough have been collected. Here, symmetric key encryption methods such as Advanced Encryption Standard (AES) is used to provide ‘quantum security’ in the foreseeable future. Where high levels of security are required, a full one-time-pad protocol can be employed where the key size is as large as the entire message to be securely transmitted. After keys obtained throughout last week is used, each party would remove all the keys from its trusted key store due to the security requirement that a key cannot be reused.  
The nodes we considered are assigned normalized weights whose values are proportional to a hypothetical network traffic (i.e., data traffic between a node and all other nodes relative to the sum of that in a network). Fig.~\ref{Fig8straties} shows the resulting keys delivered under each strategy. Here, we have the following three observations from the results: 

(1) Employing S-GD can improve the total number of the secure keys beyond that achieved with S-PD or S-TD. However, a ground- to-ground communication between two nodes is encrypted by the shared key, which is from the XOR hash of the secure keys of one node to any other destination node. In other words, the utilization of secure keys is the proportion of the resulting number of shared keys determined by a traffic distribution to that of the secure keys. In fact, consider that their one-week availability, though more secure keys delivered to Urumqi or Lhasa, delivering secure keys that exceeds individual needs may has no improvement on the utilization of keys. Nonetheless, as shown in Fig.~\ref{Fig8straties}, the secure keys about 7,468 kbits are delivered to the node of Urumqi, which embodied the same keys as that of the fibre-based QKD that relayed at least 5 ideal ground nodes to Xi’an. This indicates that S-GD has its distinct advantages in aspect of achieving a long-distance QKD.

(2) Employing S-PD promotes more secure keys being delivered to higher-weighted nodes, while the keys may cannot be promised to be delivered to lower-priority nodes. For instance, for the node, Shanghai, the number of secure keys under S-PD is 8,332 kbits, achieving a 2,551 kbit gain than S-GD, while no keys were delivered to Hefei and Jinan. This is due to the iteration of finding an optimal solution. Specifically, nodes with a lower priority, such as Hefei and Jinan, are rarely chosen in cases where a higher-priority node can access. In addition, the total number of keys delivered under S-PD is lower than that under S-GD since a node under the chosen priority level is suboptimal in terms of the link loss. Therefore, S-PD is more suitable for different levels of encryption missions.

(3) Employing S-TD achieves a flexible schedule, which makes the delivery distribution coincident with the expectation to guarantee individual needs. In Fig..~\ref{Fig9proportion}, we verify the superiority of S-TD, whose delivered proportion achieves consistent distribution with that of the expected distribution according to the weights. We remark that, clearly, although the total number of secure keys delivered under S-TD is lower than others, the consistent distribution indicates the best performance in preserving individual needs, which can further promote the utilization of the delivered secure keys.

\section*{DISCUSSION}
In this work, we develop a comprehensive framework for analysing the performance of the hypothetical but possible network consists of Micius satellite and 11 locations of China, which is a possible paradigm in future satellite-based quantum network. Our comprehensive framework integrated with orbital modelling and Himawari-8 cloud cover data statistics was proposed to enable an accurate pre-assessment of SatQKD. In the dynamic satellite-to-ground channel modelling, the beam diffractions and atmosphere effects as function of the propagation distance and elevation angle. The cloud cover contribution to channel transmission disturbance is incorporated with different local meteorological data to set up a quasi-experimental scenario. We provide a trade-off consideration between the duration available for SatQKD and its corresponding average link loss during one orbit, which could be used to support decisions involving the locations of nodes and the selection of the orbital parameters for a satellite. Lastly, we demonstrate that the use of schedule is an effective way in networking a communication system of OGSs. By formulating a problem that considers both the individual needs of nodes, we designed three different optimization methods to find the optimal solution for different scheduling strategies.

However, SatQKD operating under these scheduling strategies can in certain cases outperform fibre-based QKD with relays, with the drawbacks that a difference between the delivered distribution and the expected distribution emerges and that the mismatch limits the XOR hash operations and thus reduces the utilization of the delivered secure keys. To address this, we considered the S-TD strategy and introduced KL divergence to the algorithm to make the delivery distribution coincident with the expectation. As a result, we found that the benefits of the S-TD strategy are apparent in terms of ensuring individual needs and even promoting the utilization of the delivered keys in practical use. We explored the possibility of using satellites with various orbital heights to perform the proposed schedule. Contrary to expectations of an improvement in the total number of keys delivered by an increase in orbit altitude to increase the available duration, the results demonstrated that mitigating the geometric loss is the first consideration in future selection of the satellite orbit.

There is no denying that a satellite operating as the ‘trusted node’ to distribute keys would be replaced as an ‘untrusted node’ in the future term, with the improved fidelity of entanglement sources or possible uplink. Nonetheless, once the schedules are employed, the current configuration could show higher flexibility than the entanglement-based QKD scheme because the latter requires that two downlinks are feasible simultaneously. Moreover, a network traffic graph is also needed to distribute the pairwise keys when employing entanglement-based QKD. Notably, these two schemes, in fact, are not in conflict and even can be integrated together for different missions. This allows us to make full use of the limited time to distribute more keys. Given the increasingly complex situation, SatQKD applications such as their schedule design will be urgently needed. Conveniently, our proposed framework may be used to explore this issue in future work.

In summary, our work not only provides a practical solution to achieve network encryption without the need for fibre-based relays but can also be used as a pathfinder to support decisions involved in the selection of future quantum communication.


\begin{backmatter}

\section*{Acknowledgements}
We gratefully thank Stefano Pirandola, Xuan Han for helpful discussions.

\section*{Funding}
This work is supported by the National Natural Science Foundation of China (Grants No. 61971436); the National University of Defense Technology (19-QNCXJ-009).

\section*{Abbreviations}
QKD, Quantum Key Distribution; QUESS, Quantum Experimental Science Satellite; SatQKD, satellite-based QKD; LEO, Low Earth Orbit; OGS, Optical Ground Station; MDI, Measurement Device Independent; GEO, Geosynchronous Orbit; TLE, Two-line Elements; LLA, Latitude-Longitude-Altitude; BB84, Bennett \& Brassard 1984 QKD protocol; GLLP, Gottesman–Lo–Lütkenhaus–Preskill; QBER, Quantum Bit Error Rate; MIP, Mixed-Integer Program; KL, Kullback-Leibler; UTC, Universal Time Coordinated; RAAN, Right Ascension of Ascending Node; AES, Advanced Encryption Standard.

\section*{Availability of data and materials}
The code that contributed to the results of this study is available on request from the corresponding authors.

\section*{Ethics approval and consent to participate}
Not applicable.

\section*{Competing interests}
The authors declare that they have no competing interests.

\section*{Authors' contributions}
Xingyu Wang performed the bulk of this work at National University of Défense Technology during an exchange stay. Major of SatQKD modelling framework was worked by Xingyu Wang in this study. Jiahua Wei and Tianyi Wu designed schedule optimizations. Lei Shi modified the manuscript. With thanks to Chen Dong and Tianyi Wu from National University of Défense Technology who mooted the concept of possible quantum satellite applications.


\bibliographystyle{bmc-mathphys} 
\bibliography{bmc_article}      




\section*{Figures}
\begin{figure}[h!]
\includegraphics[width=0.9\textwidth]{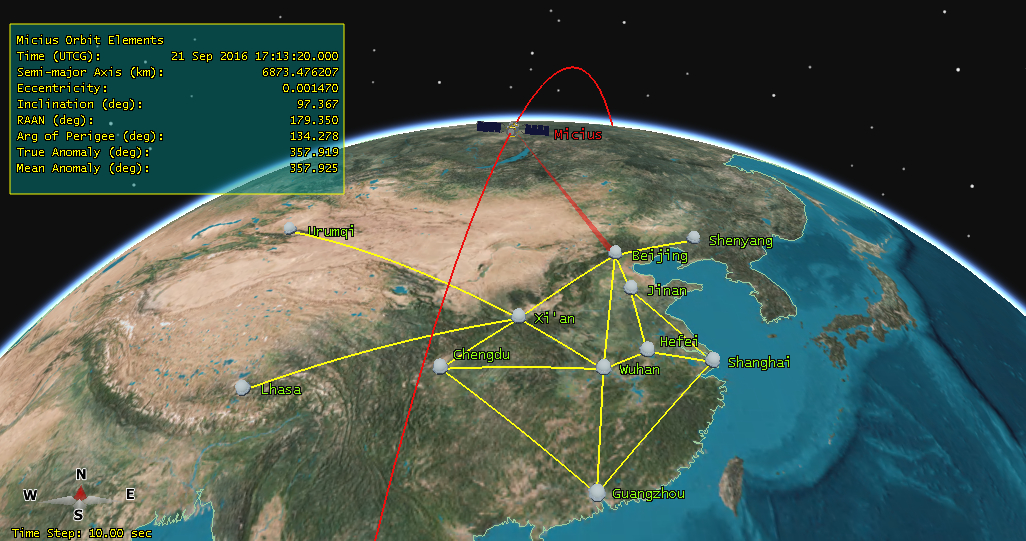}\\
\caption{An illustration of a hypothetical but possible integrated space-to-ground quantum communication network. The network consists of a space-based mobile platform (i.e., Micius) and 11 selected locations of China with OGSs: Urumqi, Lhasa, Xi’an, Chengdu, Shenyang, Beijing, Jinan, Hefei, Wuhan, Shanghai, and Guangzhou. Here, the encrypted communication in such a network is assisted by several backbone fibre links (yellow lines), using the satellite as the only trusted relay. In particular, the satellite carries out QKD operations with distinct OGSs to establish independent keys with each of them, and it subsequently broadcasts the XOR hash of both keys over a public channel, allowing any two nodes to have a shared key. In a near term, satellite only has one source on board, which can perform SatQKD to only one OGS at a time. These limitations hence suggest that, to improve the SatQKD efficiency, an efficient scheduling of downlink to the set of OGSs is required.}
\label{Fig1illustration}
\end{figure}

\begin{figure}[h!]
\includegraphics[width=0.9\textwidth]{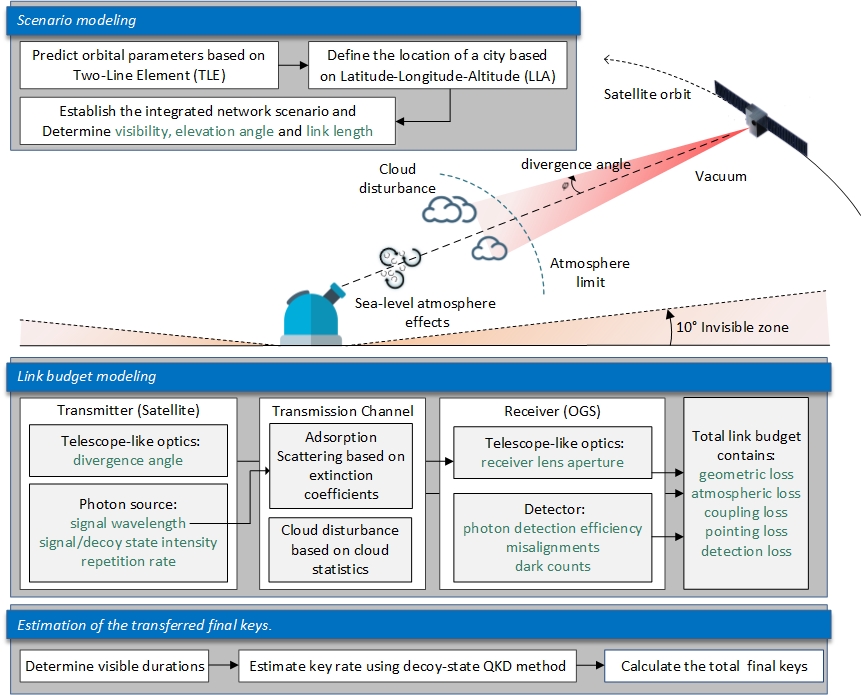}\\
\caption{The framework for modelling space-based QKD. First, based on the data file format referred to TLE sets, an exact orbital element of a specific satellite together with geographic locations of a given city can be used as inputs to determine the visibility, elevation angle and relative distance. Then, link loss budget is estimated by combining the QKD experimental transceiver parameters and cloud cover statistics involving three procedures relevant for optical signal attenuation: sending, transmission, and receiving in SatQKD system. Using the decoy-state QKD to estimate key rate per second, the total keys in a determined interval can be calculated as their result sum.}
\label{Fig2framework}
\end{figure}

\begin{figure}[h!]
\includegraphics[width=0.9\textwidth]{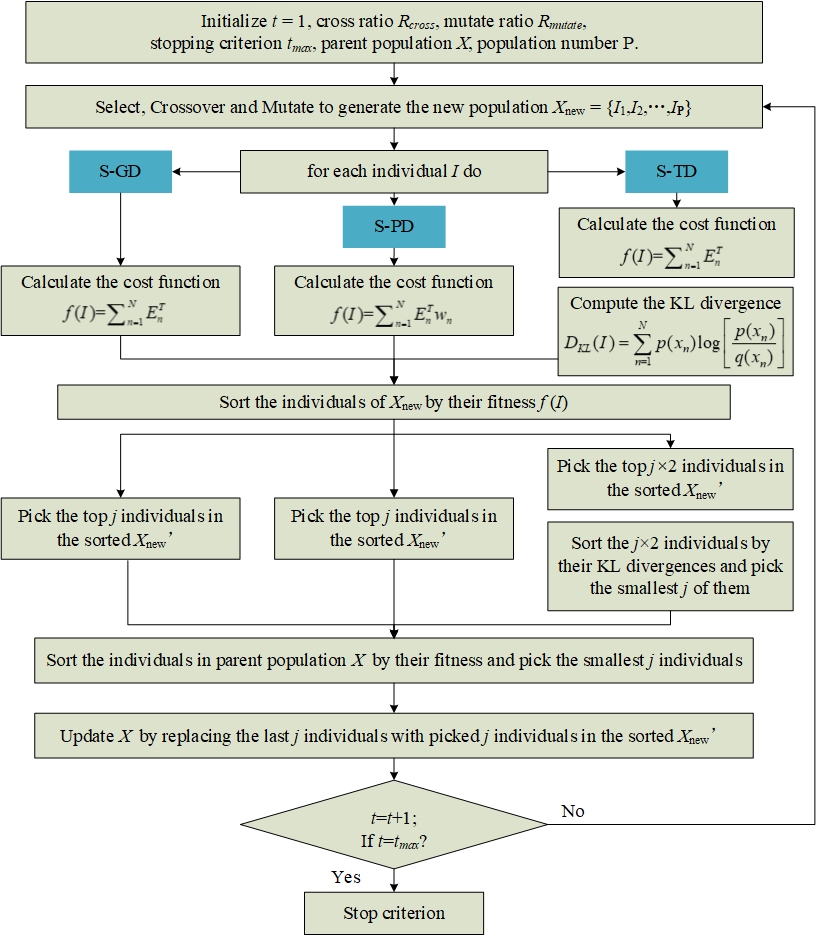}\\
\caption{Code snippets of the Algorithms 1–3.}
\label{Fig3Algorithms}
\end{figure}

\begin{figure}[h!]
\includegraphics[width=0.9\textwidth]{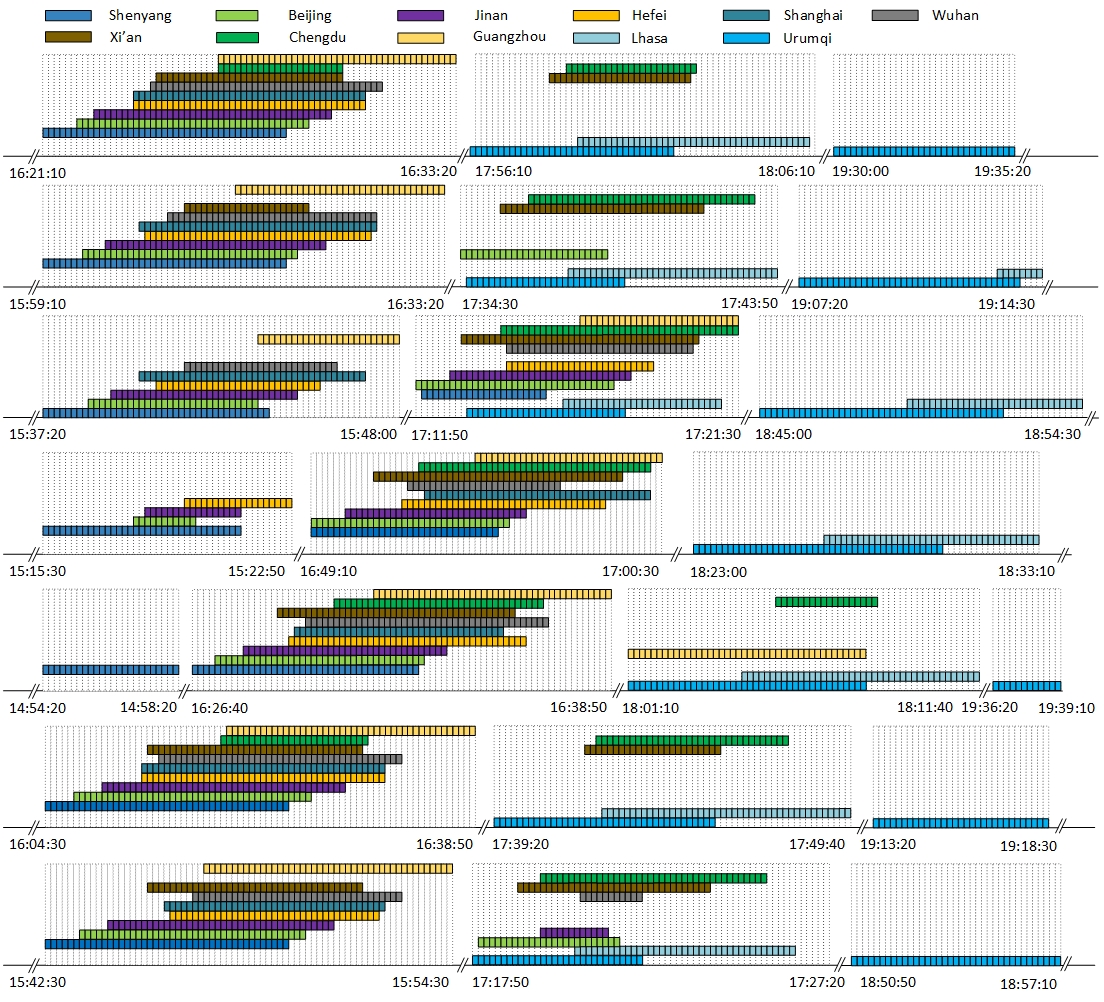}\\
\caption{Duration distribution. Illustration of the timeline, where the nodes that access to the Micius satellite in each interval (10 seconds) are labelled with different colours.}
\label{Fig4Duration}
\end{figure}

\begin{figure}[h!]
\includegraphics[width=0.46\textwidth]{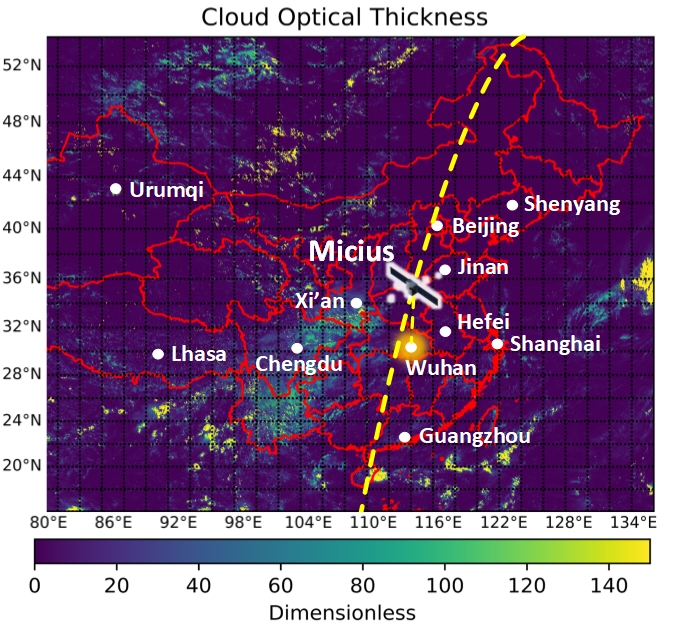}
\includegraphics[width=0.46\textwidth]{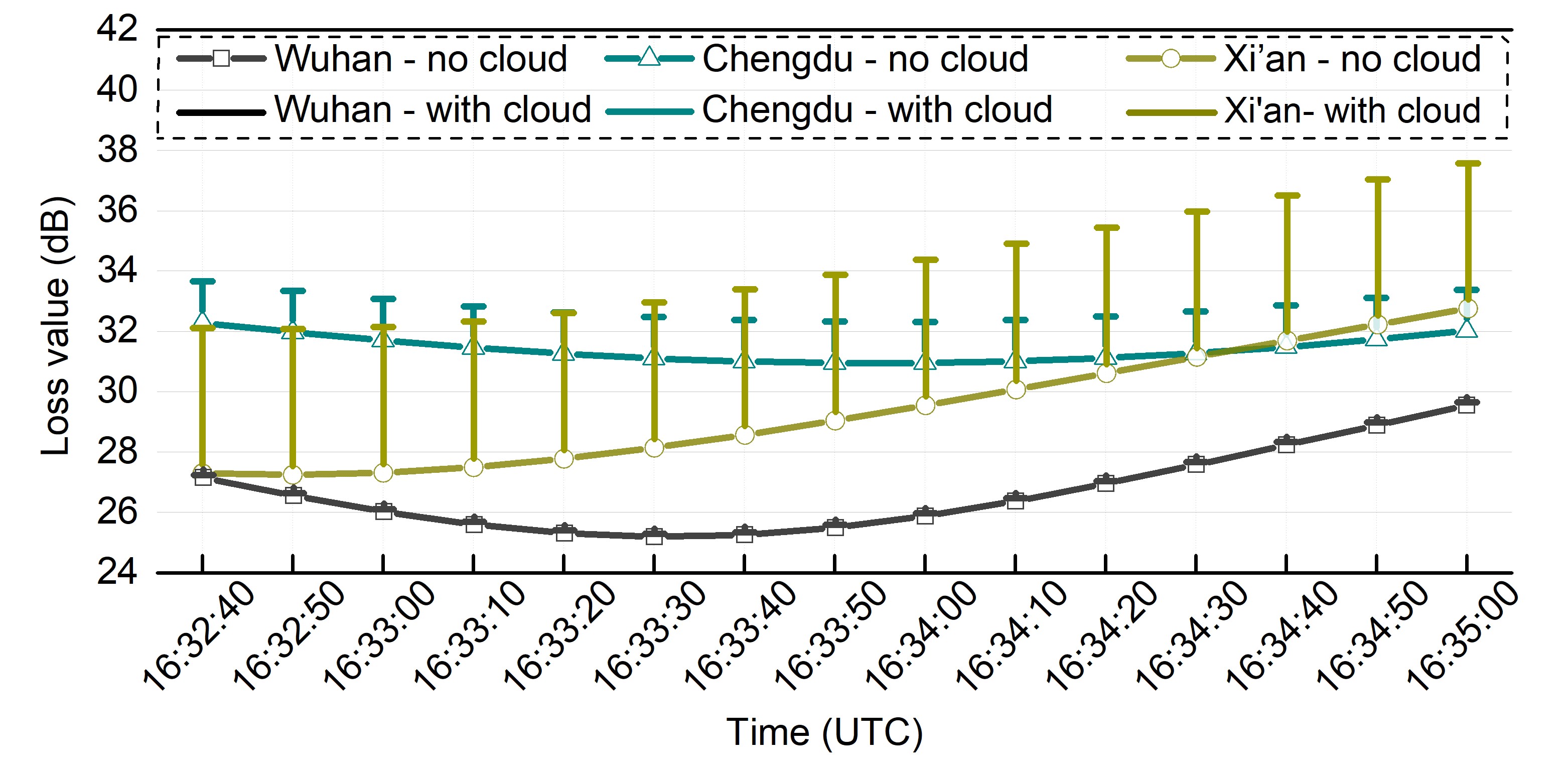}\\
\caption{Cloud cover contribution to the transmission disturbance. (a) Cloud cover distribution. The time of cloud distribution over China is Sep 23rd 2016, 16:32:40 UTC (b) Variations in the total loss budget during a common-visible interval of the nodes, Wuhan, Chengdu, and Xi’an.}
\label{Fig5Cloudcover}
\end{figure}

\begin{figure}[h!]
\includegraphics[width=0.45\textwidth]{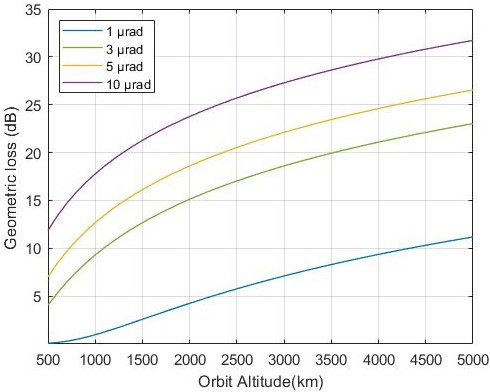}\\
\caption{Geometric loss vs. satellite altitude.  Trends of the geometric loss as a function of the satellite altitude for several values of the divergence angle.}
\label{Fig6Gloss}
\end{figure}

\begin{figure}[h!]
\includegraphics[width=0.95\textwidth]{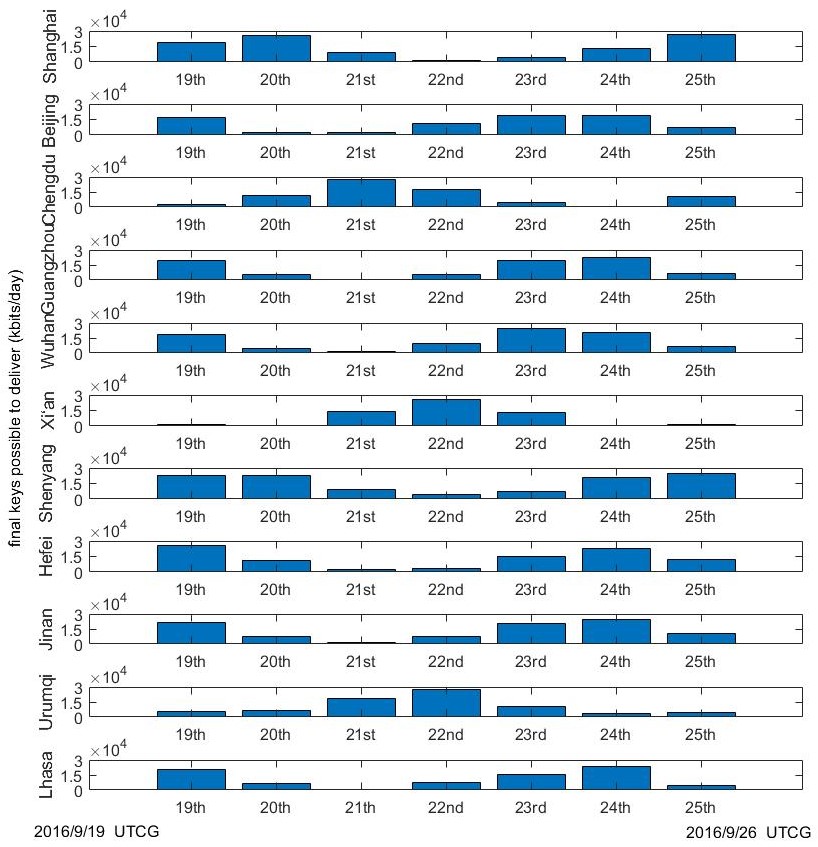}\\
\caption{The expected results of the secure keys could be delivered to the ground nodes during the week of 19th Sep 2016. For simplicity, we assume that the total key is the sum of the estimated asymptotic key rate per second during the week. Kindly note that, the finite-size keys can be considered using the security analysis of QKD with small block length~\cite{Lim2021,Thomas2022}.}
\label{Fig7keys}
\end{figure}

\begin{figure}[h!]
\includegraphics[width=0.95\textwidth]{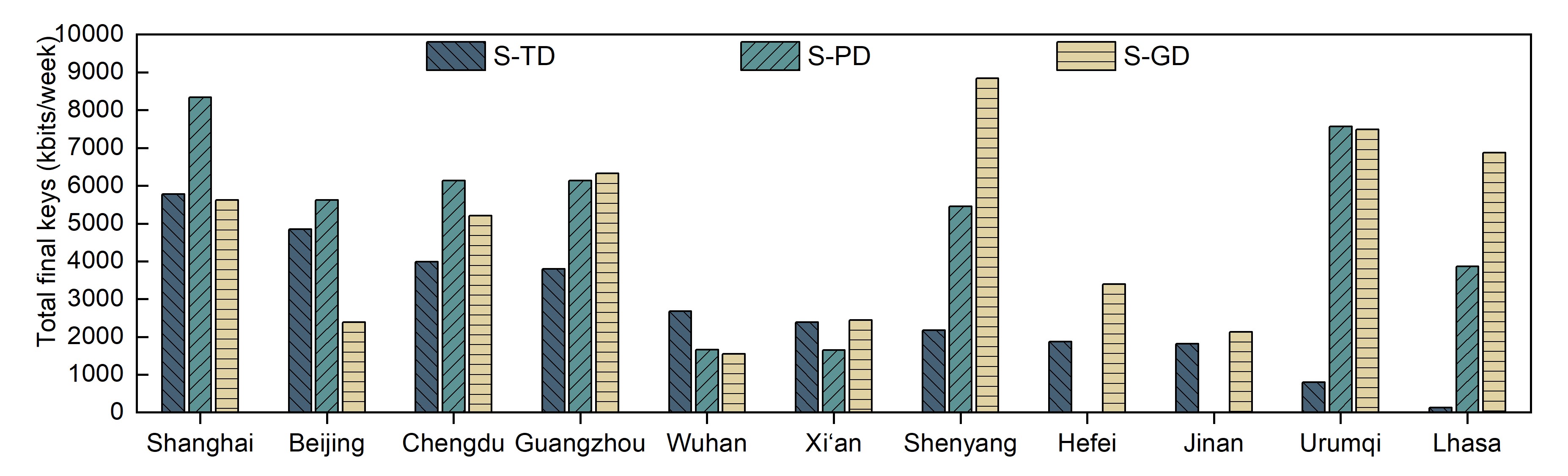}\\
\caption{Comparison of the number of secure keys under different scheduling strategies. The number of final keys delivered to the nodes under different strategies: pursuing a distribution of final keys delivered that is consistent with that of the data traffic (dark green); prioritized delivery to higher-weighted nodes (light green); pursuing only the maximization of the total number of final keys (yellow).}
\label{Fig8straties}
\end{figure}

\begin{figure}[h!]
\includegraphics[width=0.8\textwidth]{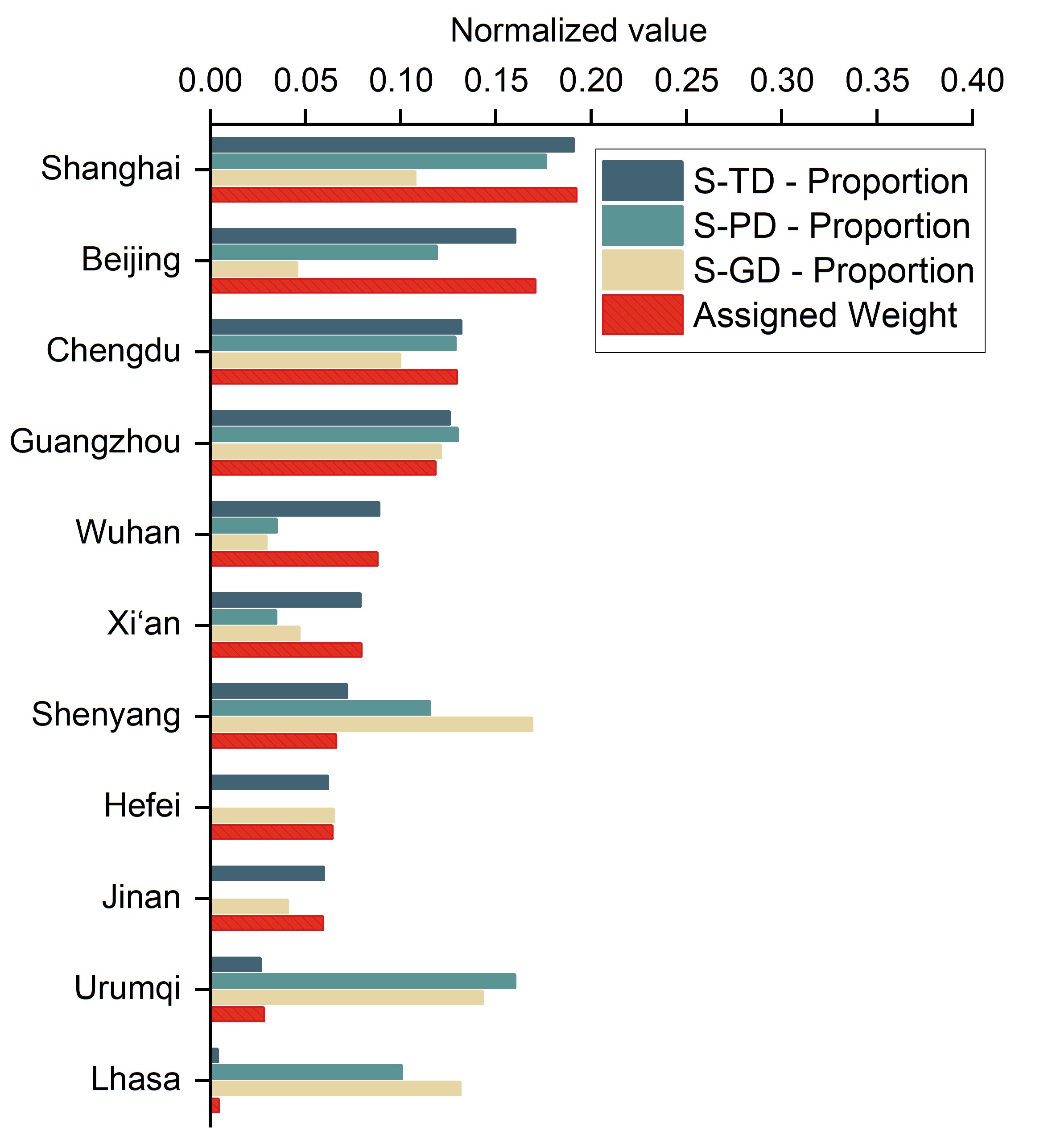}\\
\caption{Comparison of the proportion distributions. For convenience, the values of the assigned weights are temporarily proportional to the population in each city. These values also can be modified by a practical network traffic statistic.}
\label{Fig9proportion}
\end{figure}

\section*{Tables}
\begin{table}[h!]
\caption{Simulation parameters. Note that the background rate has been chosen as a fixed value while varies per second. Modelling the dynamic background photon rate requires the decay of solar irradiance, and so on~\cite{Khatri2021,PhysRevResearch.3.013279}. Work along these lines will be our future work.}
\label{Tab1}
\begin{tabular}{ll}
	\hline
	Parameters  & Value\\ \hline
	Divergence angle ($\rm{\mu rad}$)	& 10\\
	Signal wavelength (nm)	&1550\\
	Fixed signal state intensity 	&0.5\\
	Fixed decoy state intensity 	&0.08\\
	Repetition rate of the source (MHz)	&200\\
	extinction coefficient (dB)	&2\\
	Fixed pointing efficiency (dB)	&2\\
	Fixed coupling efficiency (dB)	&3\\
	Detection efficiency (dB)	&3\\
	Receiver lens aperture (m)	&1.2\\
	Error probability of dark counts	&0.5\\
	Error probability of optical misalignment	&0.015\\
	Error-correction efficiency	&1.16\\
	Fixed background rate	&${\rm{3}} \times {\rm{1}}{{\rm{0}}^{{\rm{ - 6}}}}$\\
	\hline
\end{tabular}
\end{table}

\begin{table}[htbp]
	\centering
	\scriptsize
	\caption{Variations on link budget over the visible duration of the cities for satellites at different orbit types and diffraction angles of satellite-based transmitting telescope.}
	\begin{adjustwidth}{-1.3in}{0in} 
		\setlength{\tabcolsep}{0.6mm}{
			\begin{tabular}{m{1.2cm}<{\centering}m{1.5cm}<{\centering}m{1.5cm}<{\centering}lllllllllll}
				\toprule
				\multirow{2}[2]{*}{\textbf{Orbit type}} & \multirow{2}[2]{*}{\makecell[c]{\textbf{Total visible}\\ \textbf{duration} \textbf{ (s) }}} & \multirow{2}[2]{*}{\makecell[c]{\textbf{Diffraction}\\\textbf{angle (urad)}}} & \multicolumn{11}{c}{\textbf{Variations on link budget (dB)}} \\
				\cmidrule{4-14}        &     &     & \textbf{Shanghai} & \textbf{Beijing} & \textbf{Chengdu} & \textbf{Guangzhou} & \textbf{Wuhan} & \textbf{Xi'an} & \textbf{Shenyang} & \textbf{Hefei} & \textbf{Jinan} & \textbf{Urumqi} & \textbf{Lhasa} \\
				\midrule
				\multirow{4}[1]{*}{\makecell[l]{LEO \\($\sim$500km)}} & \multirow{4}[1]{*}{12240} & 1   & 9.32-14.28 & 9.65-15.53 & 9.27-14.51 & 9.42-14.09 & 9.36-14.02 & 9.48-19.9 & 9.4-15.19 & 9.46-14.25 & 9.36-14.32 & 9.3-17.71 & 9.35-17.11 \\
				&     & 3   & 11.52-22.28 & 12.58-23.52 & 11.47-22.45 & 12.08-22.07 & 11.8-22.02 & 11.65-27.9 & 11.77-23.22 & 11.74-22.24 & 11.79-22.33 & 11.5-25.71 & 11.92-25.03 \\
				&     & 5   & 14.79-26.61 & 15.93-27.84 & 14.75-26.77 & 15.53-26.39 & 15.17-26.35 & 14.92-32.22 & 15.09-27.55 & 15.05-26.56 & 15.15-26.65 & 14.78-30.03 & 15.34-29.34 \\
				&     & 10  & 20.29-32.58 & 21.45-33.81 & 20.25-32.74 & 21.11-32.36 & 20.71-32.32 & 20.41-38.19 & 20.62-33.52 & 20.56-32.53 & 20.69-32.62 & 20.28-36.01 & 20.91-35.31 \\
				\multirow{4}[0]{*}{\makecell[l]{MEO \\($\sim$2500km)}} & \multirow{4}[0]{*}{45370} & 1   & 15.22-Inf & 15.33-25.98 & 14.94-23.07 & 15-22.51 & 15.04-23.11 & 15.12-42.37 & 15.12-26.79 & 15.12-22.84 & 15.15-26.29 & 15.11-Inf & 14.95-Inf \\
				&     & 3   & 24.04-Inf & 24.17-35.34 & 23.74-32.43 & 23.8-31.87 & 23.84-32.47 & 23.93-51.73 & 23.93-36.15 & 23.92-32.2 & 23.95-35.65 & 23.92-Inf & 23.75-Inf \\
				&     & 5   & 28.42-Inf & 28.56-39.77 & 28.12-36.86 & 28.18-36.29 & 28.23-36.9 & 28.31-56.16 & 28.31-40.57 & 28.3-36.63 & 28.34-40.08 & 28.31-Inf & 28.14-Inf \\
				&     & 10  & 34.42-Inf & 34.56-45.78 & 34.12-42.87 & 34.18-42.31 & 34.23-42.91 & 34.31-62.17 & 34.31-46.59 & 34.3-42.64 & 34.34-46.09 & 34.31-Inf & 34.14-Inf \\
				\multirow{4}[0]{*}{\makecell[l]{MEO \\($\sim$5000km)}} & \multirow{4}[0]{*}{81910} & 1   & 21.24-Inf & 21.51-30.39 & 21.25-27.21 & 21.27-27.24 & 21.24-28.92 & 21.62-Inf & 21.28-Inf & 21.32-27.68 & 21.21-27.64 & 21.24-Inf & 21.15-Inf \\
				&     & 3   & 30.62-Inf & 30.89-39.87 & 30.63-36.69 & 30.65-36.72 & 30.62-38.4 & 31-Inf & 30.66-Inf & 30.71-37.16 & 30.59-37.12 & 30.62-Inf & 30.54-Inf \\
				&     & 5   & 35.05-Inf & 35.32-44.31 & 35.06-41.12 & 35.08-41.15 & 35.05-42.83 & 35.42-Inf & 35.09-Inf & 35.14-41.59 & 35.01-41.55 & 35.05-Inf & 34.96-Inf \\
				&     & 10  & 41.06-inf & 41.34-50.33 & 41.08-47.14 & 41.09-47.17 & 41.07-48.85 & 41.44-Inf & 41.11-Inf & 41.15-47.61 & 41.03-47.57 & 41.06-Inf & 40.98-Inf \\
				\multirow{4}[1]{*}{\makecell[l]{GEO\\ ($\sim$35863km)}} & \multirow{4}[1]{*}{358700} & 1   & 36.61-Inf & 36.82-Inf & 36.68-42.45 & 36.46-Inf & 36.59-42.38 & 36.69-Inf & 36.87-Inf & 36.62-Inf & 36.73-42.52 & 37.13-Inf & 36.73-Inf \\
				&     & 3   & $>$45 & $>$45 & $>$45 & $>$45 & $>$45 & $>$45 & $>$45 & $>$45 & $>$45 & $>$45 & $>$45 \\
				&     & 5   & $>$50 & $>$50 & $>$50 & $>$50 & $>$50 & $>$50 & $>$50 & $>$50 & $>$50 & $>$50 & $>$50 \\
				&     & 10  & $>$55 & $>$55 & $>$55 & $>$55 & $>$55 & $>$55 & $>$55 & $>$55 & $>$55 & $>$55 & $>$55 \\
				\bottomrule
				\multirow{2}[2]{*}{\makecell[l]{Apart from geometric attenuation and atmospheric attenuation, single photon detection efficiency  (3 dB), pointing efficency (2 dB)  and coupling efficiency (3 dB) are included in \\the total link budget. Moreover, consider the situantion where the transmission is completely obstructed by cloud cover so that the vaule of budget would be 'Inf'.}}
		\end{tabular}}
	\end{adjustwidth}
	\label{Tab2}%
\end{table}%

\end{backmatter}
\end{document}